# Simultaneous observation of concurrent two-dimensional carbon and chlorine/bromine isotope fractionations of halogenated organic compounds on gas chromatography

**Caiming Tang[a],*, Jianhua Tan[b]**

[a] *State Key Laboratory of Organic Geochemistry, Guangzhou Institute of Geochemistry, Chinese Academy of Sciences, Guangzhou 510640, China*

[b] *Guangzhou Quality Supervision and Testing Institute, Guangzhou, 510110, China*

*Corresponding Author.

Tel: +86-020-85291489; Fax: +86-020-85290009; E-mail: CaimingTang@gig.ac.cn (C. Tang).

## ABSTRACT

It has been reported that isotope fractionation can occur on gas chromatography (GC), yet little is known about concurrent dual-elements isotope fractionations on GC. Revelation of concurrent two-dimensional carbon and chlorine/bromine isotope fractionations of halogenated organic compounds (HOCs) on GC may be of important significance for compound-specific isotope analysis (CSIA). This study presents an in-depth investigation of the two-dimensional C and Cl/Br isotope fractionations of HOCs on GC using GC-double focus magnetic-sector high resolution mass spectrometry (GC-DFS-HRMS). The two-dimensional C and Cl/Br isotope fractionations of four organochlorines and four bromobenzenes on GC were simultaneously measured by GC-DFS-HRMS. The isotope fractionations were evaluated with isotope ratios, relative variations of isotope ratios ($\Delta^h E$) and isotope fractionation extents ($\Lambda^h E$). All the HOCs exhibited significant inverse C and Cl/Br isotope fractionations, with $\Lambda^{13}C$ of 38.14‰-307.56‰, $\Lambda^{37}Cl$ of 59.60‰-146.85‰, and $\Lambda^{81}Br$ of 25.89‰-142.10‰. The isotope fractionations were significant in both ends of chromatographic peaks, while the isotope ratios in center retention-time segments were the closest to comprehensive isotope ratios in the whole peaks. Significant correlations between C isotope fractionation and Cl/Br isotope fractionation were observed, indicating that the isotope fractionations might have strong relationships and/or be dominated by similar factors. Relevant mechanisms for the two-dimensional C and Cl/Br isotope fractionations were tentatively proposed on basis of a modified two-film model and the theories related to zero point energy. The results of this study gains new insights into concurrent two-dimensional isotope fractionation behaviors of HOCs during physical processes, and are conducive to CSIA studies involving C, Cl and Br for obtaining high-quality data, particularly to dual-elements CSIA of C and Cl/Br.

## 1. Introduction

Gas chromatography (GC) separation is a critical process that can significantly influence data quality in compound-specific isotope analysis (CSIA) using GC offline or on-line isotope ratio mass spectrometry (IRMS) [1-6]. Due to its powerful separation capability, GC can separate molecules with highly similar structures such as various isomers [7-11], enantiomers [12-14] and diastereoisomers [15], and even structurally identical isotopologues and isotopomers, for instance, hydrogen vs. deuterium/tritium isotopologues/isotopomers [16-21].

Chromatographic separation of isotopes manifests as retention-time shifts between the isotopes, which is also called isotope chromatography [22-24]. Separation of isotopologues on GC is a type of isotope fractionation caused by physical processes including partition, dissolution, volatilization and diffusion [25,26]. The phenomena, mechanisms, and applications of the hydrogen isotope fractionation on GC have been well studied previously [16,17,27,28]. Generally, GC separation comprises two isotope-sensitive processes, i.e., condensation-vaporization of analytes, and diffusion of analytes with carrier gas [29,30]. The former process makes a dominant contribution to the inverse isotope effects, due to that intermolecular Van der Waals forces take effect in condensed phase and thus result in variations of isotope-sensitive zero point energy (ZPE) between condensed and gaseous states of analyte molecules [29]. To the contrary, the latter process may contribute predominantly to the normal isotope effects [30].

Up to date, most studies involving isotope fractionation on GC focused on investigating the hydrogen and carbon isotope effects through utilization of artificially isotope-labeled standards [20,29,31,32]. Only very limited studies reported the observation and mechanisms of chlorine and/or bromine isotope fractionations on GC [2,30], and very few reports concerned the carbon isotope fractionation on GC for compounds with carbon isotope ratios at the natural scale [1]. Previous studies have revealed that carbon isotope fractionation of organic compounds can take place on GC [1,33], so do chlorine and bromine isotope fractionations [2,30]. Accordingly, it can be anticipated that concurrent carbon and chlorine/bromine isotope fractionations of halogenated organic compounds (HOCs) are able to occur during GC separation. At present, however, no study has reported simultaneous observation of two-dimensional carbon and chlorine/bromine isotope fractionations of HOCs on GC. On the other hand, owing to the crucial significances of dual-elements isotope effects in environmental research, dual-elements isotope analysis has been raising scientific concerns [33-36]. Exploring the concurrent two-dimensional carbon and chlorine/bromine isotope fractionations on GC may provide insights into the method development of dual-elements isotope analysis of carbon and chlorine/bromine, and could benefit the studies involving concurrent dual-elements isotope effects of HOCs, particularly for the isotope effects triggered by physical changes. Thus, the two-dimensional carbon and chlorine/bromine isotope fractionations on GC merit in-depth investigation in terms of actual isotope fractionation behaviors, possible mechanisms and prospective applications in isotope analysis studies.

In our previous study, we have revealed the observation and probable mechanisms of chlorine/bromine isotope fractionation of HOCs on GC [30]. On the basis of our previous findings, we conducted a further study concerning the concurrent two-dimensional carbon and chlorine/bromine isotope fractionations on GC using GC-double focus magnetic-sector high resolution MS (GC-DFS-HRMS). The two-dimensional isotope fractionations were simultaneous measured by GC-HRMS, and the relevant mechanisms were tentatively proposed. The results obtained in this study may be beneficial for CSIA studies involving carbon, chlorine and bromine, particularly for dual-elements isotope analysis of carbon and chlorine/bromine of HOCs.

## 2. Experimental section

*2.1. Chemicals and materials*

Stock standard solutions of 2,2',5-trichlorobiphenyl (PCB-18), 2,4,4'-trichlorobiphenyl (PCB-28), 2,2',5,5'-tetrachlorobiphenyl (PCB-52), monobromobenzene (MoBB), 1,3,5-tribromobenzene (TrBB), pentabromobenzene (PeBB), hexabromobenzene (HBB) were purchased from Accustandard Inc. (New Haven, CT, USA). Methyl-triclosan (99.5%, Me-TCS) were obtained from Dr. Ehrenstorfer (Augsburg, Germany). Full names, structures, and other relevant information of the chemicals are listed in Table S-1.

Isooctane and nonane were of chromatographic grade and purchased from CNW Technologies GmbH (Düsseldorf, Germany) and Alfa Aesar Company (Ward Hill, MA, USA), respectively. Reference standard perfluorotributylamine used for HRMS calibration was obtained from Sigma-Aldrich LLC. (St. Louis, MO, USA).

*2.2. Stock and working solutions*

The purchased stock standards except Me-TCS were in form of either mixed or individual solutions with solvents including isooctane, toluene, acetone and methanol. Powder standard Me-TCS was weighed and dissolved in isooctane to prepare a stock solution at 1.0 mg/mL. All stock standard solutions were further diluted with nonane or isooctane to get working solutions at the concentrations suitable for GC-HRMS analysis (Table S-1). All the standard solutions were stored in a refrigerator at -20 ºC.

*2.3. Instrumental analysis*

The GC-HRMS system comprised dual Trace-GC-Ultra gas chromatographs coupled with a DFS-HRMS and a TriPlus auto-sampler (GC-DFS-HRMS, Thermo-Fisher Scientific, Bremen, Germany). The working standard solutions were directly injected onto the GC-HRMS. A DB-5MS capillary column GC column (60 m × 0.25 mm, 0.25 µm thickness, J&W Scientific, USA) was used for chromatographic separation. The carrier gas was helium at a constant flow rate of 1.0 mL/min. The inlet port and transfer line were set at 260 ºC and 280 ºC, respectively. In addition, three GC temperature programs were applied to the separation of different categories of compounds. Details of the temperature programs are provided in Table S-1.

The working conditions of the HRMS are documented as follows: electron ionization source in positive mode (EI+) was used; EI energy was kept at 45 eV; temperature of ionization source was set at 250 ºC; filament current was maintained at 0.8 mA; multiple ion detection (MID) mode was applied; dwell time was 20±2 ms for one isotopologue ion; mass resolution (5% peak-valley definition) was higher than 10000 and the MS detection accuracy was ±1 mu. The HRMS was calibrated with perfluorotributylamine in real time during MID operation.

Chemical structures of the investigated compounds were depicted by ChemDraw (Ultra 7.0, Cambridgesoft), and the exact molecular masses of the isotopologues were calculated with mass accuracy of 0.00001 u. Only the chlorine/bromine isotopologues containing none or only one $^{13}C$ atom were involved. For a molecule containing *n* Cl/Br atoms and none/one $^{13}C$ atom, all its molecular isotopologues (*n*+1) were chosen. The mass-to-charge ratios (*m/z*) were obtained

4n/aby subtracting the relative exact mass of an electron from the exact molecular weight of individual isotopologues. The *m/z* values were then imported into the MID module of HRMS for monitoring the investigated compounds. The detailed data regarding molecular isotopologues of the investigated HOCs, including isotopologue chemical formulas, exact molecular masses and exact *m/z* values, together with the retention times are provided in Table S-2, and the representative chromatograms and mass spectra of the HOCs are shown in Figure 1.

*2.4. Data processing*

The chlorine/bromine isotope ratio (IR) was calculated with

$$IR = \frac{\sum_{i=0}^{n} i \times I_i}{\sum_{i=0}^{n} (n-i) \times I_i} \quad (1)$$

where *n* is the number of Cl/Br atoms of a molecular ion; *i* is the number of $^{37}$Cl/$^{81}$Br atoms in a molecular-ion isotopologue; $I_i$ is the MS signal intensity of the molecular-ion isotopologue *i* of which all the carbon atoms are $^{12}$C. The carbon isotope ratio was calculated by

$$IR = \frac{\sum_{i=0}^{n} I'_i}{\sum_{i=0}^{n} [m \times I_i + (m-1) \times I'_i]} \quad (2)$$

where $I'_i$ represents the MS signal intensity of the molecular-ion isotopologue *i* containing only one $^{13}$C atom (the rest carbon atoms are $^{12}$C); *m* is the number of carbon atoms of the molecular ion. As external isotopic standards of the HOCs with known carbon and chlorine/bromine isotopic compositions were unavailable, all the isotope ratios measured in this study were relative values (raw isotope ratios) without being calibrated to the scales of the Vienna Pee Dee Belemnite (VPDB) standard carbon and the standard mean ocean chlorine/bromine (SMOC/B).

Each symmetric chromatographic peak of interest in the total ion chromatogram (TIC) was divided into three segments according to its retention time range. With regard to tailing chromatographic peaks (MoBB and TrBB, Figure 1), each peak was split by two vertical lines, whose interval approximated to the peak width at half height, to get three segments, and the summit should locate in the middle segment. The MS signal intensities of the isotopologues in each segment were exported and the isotope ratios were then calculated. The isotope ratio calculated with each entire chromatographic peak was regarded as the overall isotope ratio ($IR_{overall}$). Background subtraction was performed prior to exporting MS signal by deducting signal intensities of the baseline areas adjacent to both ends of each chromatographic peak. Data from six replicated injections were applied to obtaining average isotope ratios and standard deviations (SD, 1σ).

The relative variation of isotope ratio ($\Delta^h E$) derived from a retention-time segment was calculated by

$$\Delta^h E = \left( \frac{IR_{Tj}}{IR_{overall}} - 1 \right) \times 1000\text{‰} \qquad (3)$$

where $IR_{Tj}$ represents the isotope ratio derived from the retention-time segment $j$; $^hE$ represents the heavy isotopes of interest ($^{13}C$, $^{37}Cl$ and $^{81}Br$). $\Delta^hE > 0$ and $< 0$ indicate inverse and normal isotope fractionations, respectively, while $\Delta^hE = 0$ means no isotope fractionation.

The isotope fractionation extent ($\Lambda^hE$) in a whole chromatographic peak was evaluated with

$$\Lambda^hE = \left(\frac{IR_1}{IR_3} - 1\right) \times 1000‰ \qquad (4)$$

where $IR_1$ and $IR_3$ are the isotope ratios derived from the first and the last retention-time segments (T1 and T3) of the chromatographic peak, respectively. Chlorine/bromine isotope fractionation with the $\Lambda^hE < -5‰$, within $-5‰$-$5‰$, and $> 5‰$ were regarded as normal isotope fractionation, unobservable isotope fractionation, and inverse isotope fractionation, respectively in consideration of the analytical precisions. With regard to carbon isotope fractionation, the corresponding values of $\Lambda^{13}C$ were $< -10‰$, within $-10‰$-$10‰$, and $> 10‰$, based on the relatively higher analytical uncertainties of carbon isotope ratios than those of chlorine and bromine isotope ratios. The isotope fractionations on GC can be elucidated and evaluated by isotope ratios, $\Delta^hE$ and $\Lambda^hE$.

*2.5. Method performances*

Precisions (SD) of the $IR_{overall}$ of carbon, chlorine and bromine were mostly within 0.01‰-0.05‰, 0.32‰-0.41‰ and 0.91‰-1.30‰, respectively (Table S-3), which demonstrates that the precisions of the isotope analysis method could fulfil requirements for investigating two-dimensional carbon-chlorine/bromine isotope fractionations on GC.

## 3. Results and discussion

*3.1. Measured chlorine, bromine and carbon isotope ratios*

The measured isotope ratios were the basic data for the evaluation of isotope fractionations on GC in this study. Although the isotope ratios were not calibrated with external isotopic standards, they can be applied to revealing and evaluating the isotope fractionations taking place on GC, because the isotope ratios derived from different retention-time segments of each chromatographic peak were almost synchronously measured. This study did not aim to measure the real isotope ratios but focused on exploration of the isotope fractionations occurring on GC. The isotope fractionations can be more intuitively evaluated by $\Delta^h E$ and $\Lambda^h E$, which in fact are the relative variations of isotope ratios in different retention-time segments. The $\Delta^h E$ and $\Lambda^h E$ calculated with the raw isotope ratios are certainly consistent with those calculated with the real isotope ratios calibrated with external isotopic standards (if available), due to the calculation schemes as expressed by eq 3 and eq 4. In consideration of the main objective of this study, the calibration with external isotopic standards was thus non-mandatory and unnecessary in practice. The $IR_{overall}$ of chlorine, bromine and carbon were 0.31527-0.32336 (SD ≤ 0.41‰), 0.92525-0.96303 (SD ≤ 1.30‰) and 0.00961-0.01092 (SD ≤ 0.05‰), respectively (Table S-3), demonstrating relatively high precisions of the raw isotope ratios measured by GC-HRMS.

As shown in Table S-4, the chlorine isotope ratios of PCB-18, PCB-28 and PCB-52 in the first retention-time segment (T1) were from 0.32446 (SD: 0.75‰) to 0.32671 (SD: 1.80‰), and those in the middle retention-time segment (T2) and the last segment (T3) were 0.31492-0.31630 (SD: 0.63‰-0.83‰) and 0.30581-0.30812 (SD: 0.39‰-1.29‰), respectively. The chlorine isotope ratios decreased with the retention-time segments from T1 to T3, with very similar variation tendencies among the three PCBs (Figure 2a). These chlorine isotope ratios clearly show inverse chlorine isotope fractionation of the three PCBs on GC. Similarly, Me-TCS also presented inverse chlorine isotope fractionation on GC (Figure 2b), with the chorine isotope ratios derived from T1, T2 and T3 of 0.34812 (SD: 4.58‰), 0.32342 (SD: 1.52‰) and 0.30355 (SD: 2.00‰), respectively (Table S-4). As illustrated in Figure 2a and 2b, the patterns plotted with the chlorine isotope ratios are almost straight lines for the four organochlorine compounds (Figure 2a and 2b).

The ranges of bromine isotope ratios of the investigated four bromobenzenes derived from T1, T2 and T3 were 0.94482-1.04828 (SD: 1.60‰-6.19‰), 0.92482-0.96274 (SD: 0.88‰-1.69‰) and 0. 0.91760-0.94436 (SD: 1.11‰-2.08‰), respectively (Table S-4). These data confidently show the decline of bromine isotope ratios of the bromobenzenes from T1 to T3, indicating inverse bromine isotope fractionation on GC (Figure 2c and 2d). The bromine isotope ratios derived from the three retention-time segments of PeBB and HBB form two approximately straight-line patterns, while those of MoBB and TrBB make up two folding-line patterns (Figure 2c and 2d), which may be owing to the tailing chromatographic peaks of MoBB and TrBB.

The ranges of carbon isotope ratios of all the HOCs in T1, T2 and T3 were 0.01038-0.01261 (SD: 0.02‰-0.27‰), 0.00955-0.01090 (SD: 0.03‰-0.11‰) and 0.00845-0.01035 (SD:

0.03‰-0.12‰), respectively (Table S-4). These measured isotope ratios along with their respective analysis uncertainties (SD) definitely manifest the sequential decrease of carbon isotope ratios from T1 to T3, demonstrating that the HOCs presented inverse carbon isotope fractionation on GC (Figure 3). The three PCBs showed fairly similar changing trends of the carbon isotope ratios among the three retention-time segments (Figure 3a). PeBB and HBB showed parallel carbon isotope-ratio patterns from T1 to T3 (Figure 3c). While Me-TCS, MoBB and TrBB individually showed distinctive carbon isotope-ratio patterns (Figure 3b-3d). Most HOCs showed straight-line patterns of carbon isotope ratios (Figure 3a-3c), with the exception of MoBB and TrBB, of which the isotope-ratio patterns are folding lines (Figure 3c and 3d), which may be due to the tailing chromatographic peaks of the two compounds.

*3.2. Relative variations of isotope ratios*

The measured $\Delta^h E$ ($\Delta^{37}Cl$, $\Delta^{81}Br$ and $\Delta^{13}C$) are the essential data that more straightforward and intuitively show the isotope fractionation situations than the isotope ratios (Figure 4). The $\Delta^{37}Cl$ of the three PCBs in T1, T2 and T3 were 29.15±2.60‰ to 33.71±5.65‰, −2.61±1.88‰ to 0.78±1.51‰, and −30.00±2.09‰ to −25.11±0.88‰, respectively (Table S-4). These $\Delta^{37}Cl$ in individual retention-time segments were very close, whereas those in T1 and T3 were significantly different from the corresponding $\Delta^{37}Cl$ of Me-TCS (Figure 4a), the $\Delta^{37}Cl$ of which were 76.57‰ (SD: 13.83‰), 0.19‰ (SD: 4.80‰) and −61.27‰ (SD: 6.24‰) in T1, T2 and T3, respectively (Table S-4). As illustrated in Figure 4a, the $\Delta^{37}Cl$ patterns of the three PCBs are closely adjacent and parallel, but evidently distinguishable from that of Me-TCS. This result implies that the chlorine isotope fractionation scenarios of the PCBs on GC were similar, and remarkably different form that of Me-TCS. The $\Delta^{13}$C of the three PCBs were 108.41‰-128.98‰ (SD: 7.40‰-12.36‰) in T1, −9.43‰-1.82‰ (SD: 4.04‰-7.37‰) in T2, and −121.26‰ to −92.08‰ (SD: 6.19‰-9.25‰) in T3. Me-TCS had the $\Delta^{13}$C of 155.68±22.66‰, −1.51±9.69‰ and −116.19±8.57‰ in T1, T2 and T3, respectively (Table S-4). On the other hand, the $\Delta^{13}$C patterns of all the investigated organochlorines are relatively closer and more similar compared with the $\Delta^{37}$Cl patterns (Figure 4a and 4b). This result indicates that the organochlorines exhibited relatively more consistent behaviors of carbon isotope fractionation than those of chlorine isotope fractionation.

The $\Delta^{81}$Br of PeBB and HBB in every retention-time segment were similar, with the ranges from 31.21±2.80‰ to −27.12±1.35‰, and from 22.97±3.06‰ to −18.84±1.30‰, respectively with the retention-time segments from T1 to T3 (Table S-4). These data suggest the similar bromine isotope fractionation on GC for PeBB and HBB. The $\Delta^{81}$Br patterns of MoBB and TrBB are significantly distinctive (Figure 4c). MoBB presented the $\Delta^{81}$Br in T1 (21.25±2.20‰) and T2 (−0.47±0.44‰) similar to those of PeBB and HBB (Figure 4c), while its $\Delta^{81}$Br in T3 (−4.62±1.02‰) was evidently higher than the corresponding $\Delta^{81}$Br of other bromobenzenes. On the other hand, TrBB showed the $\Delta^{81}$Br in T2 (0.64±0.760‰) and T3 (−23.29±0.74‰) similar to those of PeBB and HBB, whereas its $\Delta^{81}$Br in T1 (115.51±6.54‰) was significantly higher than the corresponding $\Delta^{81}$Br of other brominated benzenes. These findings show that MoBB and TrBB had distinctive bromine isotope fractionation scenarios on GC, possibly attributing to the peak-tailing chromatographic behaviors of the two compounds. The $\Delta^{13}$C patterns of PeBB and HBB are considerably consistent (Figure 4d), with $\Delta^{13}$C ranges in T1-T3 of 91.37±7.66‰ to −94.59±5.96‰, and 106.45±14.04‰ to −94.03±10.85‰, respectively. Like the bromine isotope fractionation scenarios, the carbon isotope fractionation situations of PeBB and HBB were similar (Figure 4d). As for MoBB and TrBB, however, the $\Delta^{13}$C patterns between the two compounds are clearly different, and also inconsistent with those of PeBB and HBB (Figure 4d). MoBB showed the lowest $\Delta^{13}$C in T1 (27.25±2.39‰), and the highest $\Delta^{13}$C in T3 (−10.49±2.09‰) among the four bromobenzenes. TrBB had the $\Delta^{13}$C in T1 (100.85±7.95‰) and $\Delta^{13}$C in T2 (2.57±1.94‰) similar to the corresponding values of PeBB and HBB, but had the significantly different $\Delta^{13}$C in T3 (−26.43±4.46‰) from that of any other brominated benzene. These observations indicate that MoBB and TrBB had respective specific carbon isotope fractionation features in contrast to other bromobenzenes.

In general, all the investigated compounds exhibited inverse carbon and chlorine/bromine isotope fractionations. The departures of $\Delta^hE$ from zero lines are large in the first and the last retention-time segments (Figure 4), which reflects that the isotope fractionations were significant in both ends of chromatographic peaks. The $\Delta^hE$ in the middle retention-time segments were close to zero and generally within the scales of analysis uncertainties, which indicates that the isotope ratios in the center retention-time segments were the closest to the IR$_{overall}$. This finding is consistent with our previous outcome [30].

*3.3. Isotope fractionation extents*

The measured $\Lambda^hE$ ($\Lambda^{37}Cl$, $\Lambda^{81}Br$ and $\Lambda^{13}C$) directly reflect the isotope fractionation magnitudes of the HOCs on GC. As shown in Figure 5, the $\Lambda^{37}Cl$ and $\Lambda^{81}Br$ were generally lower than the $\Lambda^{13}C$ for individual compounds. The three PCBs exhibited similar $\Lambda^{37}Cl$ within the range from 59.60‰ (SD: 5.22‰) to 60.98‰ (SD: 2.84‰), and Me-TCS had the highest $\Lambda^{37}Cl$ of 146.85‰ (SD: 14.70‰). MoBB, PeBB and HBB showed relatively lower $\Lambda^{81}Br$ (from 25.89±2.97‰ to 59.95±2.04‰) than TrBB, of which the $\Lambda^{81}Br$ was 142.10±6.96‰. The $\Lambda^{13}C$ of the four organochlorines were relatively similar, within the range from 235.15±9.32‰ to 307.56±17.92‰, and PCB-18 and Me-TCS had the lowest and the highest $\Lambda^{13}C$, respectively. Interestingly, we unequivocally observed a gradually ascending tendency of $\Lambda^{13}C$ with the increase of bromine atoms on the molecules (from MoBB to HBB). MoBB had the lowest $\Lambda^{13}C$ of 38.14±4.00‰, while HBB presented the highest $\Lambda^{13}C$ of 221.45±22.31‰ among the four brominated benzenes. In addition, the bromobenzenes generally showed significantly lower $\Lambda^{13}C$ in comparison with the organochlorines.

*3.4. Two-dimensional carbon and chlorine/bromine isotope fractionations*

To reveal the concurrent two-dimensional carbon and chlorine/bromine isotope fractionations of the HOCs on GC, we plotted $\Delta^{37}Cl$ vs. $\Delta^{13}C$ of the individual organochlorines and $\Delta^{81}Br$ vs. $\Delta^{13}C$ of the individual organobromine compounds, and fitted the data with linear functions (Figure 6). As Figure 6a shows, the data of all the organochlorines are fitted very well with the respective linear functions with the correlation coefficient $R^2 \geq 0.99712$, particularly for PCB-18 and PCB-28, of which the $R^2 \geq 0.99935$. The plotted lines of PCB-18, PCB-28 and PCB-52 are fairly similar, presenting similar slopes (0.24151-0.27254) and close intercepts (–0.32367-0.10688). However, the plotted line of Me-TCS is definitely different from those of other organochlorines, showing an evident higher slope of 0.51708. These data show that the three PCBs had similar characteristics of two-dimensional carbon-chlorine isotope fractionations on GC, which were distinctly different from that of Me-TCS.

In comparison with the organochlorines, the investigated organobromine compounds showed less satisfactory matching scores between the experimental data and the respective linear functions (Figure 6b), giving rise to the relatively lower correlation coefficient ($R^2$) from 0.69301 to 0.99823. However, the fitting results of the bromobenzenes are still good as the $R^2$ of TrBB, PeBB and HBB are $\geq$ 0.95726, although the correlation corresponding to MoBB is relatively less significant ($R^2 = 0.69301$). The fitted lines of the four bromobenzenes are differentiable from each other with statistical significance, which means the bromobenzenes individually had distinguishable concurrent two-dimensional carbon-bromine isotope fractionation behaviors on GC. The fitted lines of PeBB and HBB show relatively higher similarity relative to those of MoBB and TrBB. Therefore, PeBB and HBB might possess relatively more similar scenarios of two-dimensional carbon-bromine isotope fractionations, while MoBB and TrBB had more distinctive scenarios.

As can be seen from Figure 6, all the fitted lines have positive slopes, indicting the positive correlations between $\Delta^{37}Cl/\Delta^{81}Br$ and $\Delta^{13}C$, which demonstrates that the chlorine/bromine isotope fractionation was positively correlated with the carbon isotope fractionation of the HOCs on GC. This result shows that the chlorine/bromine and carbon isotope fractionations were in possession of the same direction (inverse isotope fractionation). In addition, the ratios between the isotope fractionation rates of chlorine/bromine and those of carbon for individual compounds were constant with the retention time variation. Figure 6 also shows that the slopes of all the fitted lines are less than 1, which implies that the isotope fractionation rates of chlorine/bromine were lower than those of carbon. The three PCBs, PeBB, and HBB showed low relative rates between chlorine/bromine isotope fractionation and carbon isotope fractionation with the slopes within 0.20494-0.29374, and Me-TCS together with MoBB had the medium relative rates with the slopes of 0.51708 and 0.58776, respectively while TrBB presented the highest relative rate with the slope of 0.91005. Therefore, the carbon isotope fractionation was more prominent compared with the chlorine/bromine isotope fractionation on GC for the investigated compounds.

*3.5. Tentative mechanistic interpretation*

Isotope chromatography has been studied for more than half of a century, and some explanations for the mechanisms have been proposed [16,17,25-27]. With regard to the isotope effects occurring on GC, it has been proposed that two types of contradictory isotope effects, i.e., liquid-vapor isotope effects and diffusion isotope effects, contribute together to the total (apparent) isotope effects [30]. In our precious study, we employed a modified two-film model to interpret the observed chlorine/bromine isotope effects of HOCs on GC [30]. The modified two-film model can also be applied to the mechanistic explanation for the observed concurrent two-dimensional C-Cl/Br inverse isotope fractionations of HOCs on GC in the current study. The molecules of heavier isotopologues have slightly smaller Van der Waals volumes than those of lighter isotopologues, leading to lower polarizabilities of the heavier-isotopologue molecules relative to those of the lighter-isotopologue molecules. As a consequence, the heavier-isotopologue molecules have slightly weaker intermolecular interactions with the bonded functional groups on the stationary phase materials of GC columns than those of the lighter-isotopologue ones with the bonded groups. This difference in intermolecular interactions tends to push the heavier-isotopologue molecules out from the stationary phase materials, and keep the lighter-isotopologue ones staying in the stationary phase materials. Therefore, during the liquid-vapor process (stationary phase materials are in liquid state), the heavier-isotopologue molecules escape from the stationary phase materials faster than the lighter-isotopologue ones, suggesting inverse isotope effects. Once entering into the carrier gas layer, the molecules of analytes are subjected to diffusion process, in which the diffusive effects are mass dependent—the heavier molecules fly slower than the lighter ones. Therefore, the diffusive effects drive the heavier-isotopologue molecules to diffuse more slowly than driving the lighter-isotopologue ones, showing normal isotope effects. In brief, the liquid-vapor isotope effects are inverse, whereas the diffusive isotope effects are normal. The two types of isotope effects act together to determine the magnitudes and directions of total isotope effects which can be observed as isotope fractionations on GC.

In this study, all the observed isotope fractionations were inverse, which means that all the total isotope effects were inverse and mainly dominated by the inverse liquid-vapor isotope effects.

In other words, the magnitudes of the inverse liquid-vapor isotope effects excessed those of the normal diffusion isotope effects. The observed correlations of $\Delta^{37}Cl$ vs. $\Delta^{13}C$ and $\Delta^{81}Br$ vs. $\Delta^{13}C$ were significant, indicating that the isotope fractionations of chlorine/bromine and carbon might have strong relationships, and/or were possibly controlled by similar factors. The total isotope effects between chlorine/bromine and carbon of individual HOCs on GC might maintain constant intensity ratios in the corresponding retention time ranges.

The observed carbon isotope fractionation was generally more significant than the chlorine/bromine isotope fractionation, indicating that the total isotope effects of carbon were larger than those of chlorine and bromine. According to the literatures [30,37], the total isotope effects ($IE_{total}$) can be expressed as:

$$IE_{total} = \nabla_{liq-vap} + \varepsilon_{diff-He} \tag{5}$$

where $\nabla_{liq-vap}$ is liquid-vapor isotope effects and $\varepsilon_{diff-He}$ denotes diffusion isotope effects. In addition, diffusive isotope effects can be obtained with the following equations [30,38,39]:

$$\varepsilon_{diff-He} = c(1-s)(\alpha_{diff-He} - 1) \tag{6}$$

$$\alpha_{diff-He} = \sqrt{\frac{M_l(M_h + M_{He})}{M_h(M_l + M_{He})}} \tag{7}$$

where $c$ is a correction factor for carrier gas flow ranging from 1 to 0.5, and $s$ is corresponding to the relative vapor saturation of organic compounds (less than 100%); $\alpha_{diff-He}$ represents the fractionation factors derived from the diffusive effects on GC; $M_l$, $M_h$ and $M_{He}$ are molecular weights of the lighter isotopologues, the heavier isotopologues and helium, respectively. Since $M_h > M_l$, $\alpha_{\text{diff-He}}$ are always less than 1, and $\varepsilon_{\text{diff-He}}$ are thus always negative.

We define the nominal molecular mass of a hypothesized chlorine/bromine isotopologue as $m$ (isotopologue-0), then the molecular mass of the adjacent isotopologue containing one more $^{13}C$ atom is $m+1$ (isotopologue-1), and that of the adjacent isotopologue containing one more $^{37}Cl/^{81}Br$ atom is $m+2$ (isotopologue-2) [40]. Accordingly, the diffusive fractionation factor between isotopologue-0 and isotopologue-1 ($\alpha_{\text{diff-C}}$) can be given by:

$$\alpha_{diff-C} = \sqrt{\frac{m(m+1+m_{He})}{(m+1)(m+m_{He})}} = \sqrt{\frac{m(m+5)}{(m+1)(m+4)}} \qquad (8)$$

where $m_{He}$ is the nominal molecular mass of helium (4 u), and the diffusive fractionation factor between isotopologue-0 and isotopologue-2 ($\alpha_{\text{diff-Cl/Br}}$) can be calculated with:

$$\alpha_{diff-Cl/Br} = \sqrt{\frac{m(m+2+m_{He})}{(m+2)(m+m_{He})}} = \sqrt{\frac{m(m+6)}{(m+2)(m+4)}} \qquad (9)$$

It can be readily proved that the function

$$f(m) = \sqrt{1+\frac{4}{m}} \qquad (10)$$

is monotonically decreasing in its domain of definition. Thus we have

$$\sqrt{\frac{m+5}{m+1}} > \sqrt{\frac{m+6}{m+2}} \qquad (11)$$

and further obtain $\alpha_{diff-C} > \alpha_{diff-Cl/Br}$. Therefore, according to eq 6, we have $\varepsilon_{diff-C} > \varepsilon_{diff-Cl/Br}$, where $\varepsilon_{diff-C}$ and $\varepsilon_{diff-Cl/Br}$ are the diffusion isotope effects of carbon and chlorine/bromine, respectively. Herein, it merits attention that $\varepsilon_{diff-C}$ and $\varepsilon_{diff-Cl/Br}$ are vectors instead of scalars, and the direction of inverse isotope effects is defined as positive and that of normal isotope effects is defined as negative. In fact, the absolute values of $\varepsilon_{diff-C}$ are lower than those of $\varepsilon_{diff-Cl/Br}$.

We take CCl, a chlorinated methylidyne radical [41], for an imaginary example to elucidate the liquid-vapor isotope effects. We define the chemical bonds $^{12}C-^{35}Cl$, $^{13}C-^{35}Cl$, and $^{12}C-^{37}Cl$ as bond-0, bond-1, and bond-2, respectively. According to the equations addressed in a previous study [42], we obtain the following modified equations:

$$\Delta v_1 = v_0 - v_1 = v_0 \left(1 - \sqrt{\frac{m_{35} \times m_{12}/(m_{35}+m_{12})}{m_{35} \times m_{13}/(m_{35}+m_{13})}}\right) \quad (12)$$

and

$$\Delta v_2 = v_0 - v_2 = v_0 \left(1 - \sqrt{\frac{m_{35} \times m_{12}/(m_{35}+m_{12})}{m_{37} \times m_{12}/(m_{37}+m_{12})}}\right) \quad (13)$$

where $v_0$, $v_1$, and $v_2$ are the vibrational frequencies of bond-0, bond-1, and bond-2, respectively; $\Delta v_1$ is the difference of the vibrational frequencies between $v_0$ and $v_1$, and $\Delta v_2$ is the difference of the vibrational frequencies between $v_0$ and $v_2$; $m_{12}$, $m_{13}$, $m_{35}$ and $m_{37}$ represent the atom masses of $^{12}C$, $^{13}C$, $^{35}Cl$ and $^{37}Cl$, respectively. It can be proved that $\Delta v_1 > \Delta v_2$.

In addition, according to the literature [42], we have two modified equations as follows:

$$\Delta ZPE_1 = ZPE_0 - ZPE_1 = 1/2\, h(v_0 - v_1) = 1/2\, h\Delta v_1 \qquad (14)$$

and

$$\Delta ZPE_2 = ZPE_0 - ZPE_2 = 1/2\, h(v_0 - v_2) = 1/2\, h\Delta v_2 \qquad (15)$$

where $ZPE_0$, $ZPE_1$ and $ZPE_2$, are the ZPEs of bond-0, bond-1 and bond-2 respectively; $\Delta ZPE_1$ and $\Delta ZPE_2$ are the differences of $ZPE_1$ and $ZPE_2$ relative to $ZPE_0$, respectively; $h$ is the Planck constant. As $\Delta v_1 > \Delta v_2$, we thus obtain $\Delta ZPE_1 > \Delta ZPE_2$. Therefore, the bond energy difference between bond-0 and bond-1 is slightly larger than that between bond-0 and bond-2. As a result, the bond length difference between bond-0 and bond-1 is slightly bigger than that between bond-0 and bond-2, suggesting the molecular volume variation between $^{12}C^{35}Cl$ and $^{13}C^{35}Cl$ is greater than that between $^{12}C^{35}Cl$ and $^{12}C^{37}Cl$. We accordingly conclude that substitution of a $^{12}C$ atom with a $^{13}C$ atom on an organochlorine molecule can cause larger molecular volume change relative to substitution of a $^{35}Cl$ atom with a $^{37}Cl$ atom. Then the molecular polarizability difference between the initial isotopologue (isotopologue-0, unsubstituted) and the $^{13}C$-substituted isotopologue (isotopologue-1) is larger than that between isotopologue-0 and $^{37}Cl$-substituted isotopologue (isotopologue-2). The polarizability plays a critical role in the intermolecular interactions between analyte molecules and the bonded groups on the stationary phase materials (DB-5MS GC column). The intermolecular interactions between the molecules of isotopologue-1 and the bonded groups are thus weaker than those between the molecules of isotopologue-2 and the bonded groups. The stationary phase materials thus pull the isotopologue-1 molecules with less strength than pulling the isotopologue-2 molecules, and pull the isotopologue-0 molecules with the largest strength. Therefore, the $\nabla_{\text{liq-vap}}$ between isotopologue-1 and isotopologue-0 are higher than those between isotopologue-2 and isotopologue-0. In other words, the $\nabla_{\text{liq-vap}}$ of carbon are higher than those of chlorine. Substituting chlorine with bromine, we can analogously conclude that the $\nabla_{\text{liq-vap}}$ of carbon are higher than those of bromine. In addition, as mentioned above, we have proved $\varepsilon_{\text{diff}-C} > \varepsilon_{\text{diff}-Cl/Br}$. Then, according to eq 5, we obtain $IE_{\text{total}-C} > IE_{\text{total}-Cl/Br}$, where $IE_{\text{total-C}}$ and $IE_{\text{total-Cl/Br}}$ denote the total isotope effects of carbon and chlorine/bromine, respectively. Consequently, it can be explained that the observed carbon isotope fractionation was more significant than the chlorine/bromine isotope fractionation of the HOCs on GC.

*3.6. Implications for CSIA and environmental research*

Accuracy and precision of analysis results are key factors in CSIA studies, which may be affected by instrumental uncertainties such as isotope fractionations on GC. The results in this study revealed that the carbon, chlorine and bromine isotope fractionations of analytes on GC could be very large in contrast with the isotope fractionations occurring under natural or common laboratorial conditions. Therefore, analysts should be cautious to the isotope fractionations taking place on GC, in order to reduce or eliminate negative effects on CSIA results. When preparative GC is employed to prepare samples, analytes should be collected as completely as possible, and chromatographic peaks would better to be baseline-separated. With regard to CSIA using GC-IRMS or GC coupled to other types of MS, chromatographic peaks should be separated sufficiently and integrated completely.

The isotope fractionations on GC are caused by physical processes such as dissolution, partition, condensation, vaporization and diffusion. The concurrent two-dimensional carbon and chlorine/bromine isotope fractionations on GC may able to be used as referencing scenarios of two-dimensional isotope fractionations triggered by similar physical processes in the environment or laboratory experiments. And the proposed mechanistic interpretation in this study may be helpful in elucidation of the mechanisms of two-dimensional C-Cl/Br isotope fractionations taking place during physical processes under other circumstances.

## 4. Conclusions

In this study, we systematically investigated the concurrent two-dimensional carbon and chlorine/bromine isotope fractionations of four organochlorines and four bromobenzenes on GC. The carbon and chlorine/bromine isotope fractionations were simultaneously measured by GC-HRMS, and evaluated with measured isotope ratios, $\Delta^h E$ and $\Lambda^h E$. The precisions of the isotope ratios, $\Delta^h E$ and $\Lambda^h E$ were sufficient to differentiate the modes and magnitudes of the isotope fractionations. The carbon isotope fractionation and chlorine/bromine isotope fractionation were inverse for all the HOCs, showing remarkably high $\Lambda^h E$ values. The $\Delta^h E$ deviated substantially from the zero value in the first and the last retention-time segments, while closed to the zero value in the middle segments, indicating that the isotope fractionations were significant in the both ends of chromatographic peaks and insignificant in the central regions. The modes and magnitudes of isotope fractionations were varied depending on elements and compounds. The three PCBs showed similar modes and magnitudes of carbon and chlorine isotope fractionations. The $\Delta^{37}Cl$ vs. $\Delta^{13}C$ and $\Delta^{81}Br$ vs. $\Delta^{13}C$ were plotted for revealing the internal relations of the concurrent carbon and chlorine/bromine isotope fractionations. Significant positive correlations of $\Delta^{37}Cl$ vs. $\Delta^{13}C$ and $\Delta^{81}Br$ vs. $\Delta^{13}C$ were found for all the HOCs, demonstrating that the isotope fractionations might have strong relationships and/or be dominated by similar effects. The relationships between the two-dimensional isotope fractionations varied depending on pairs of elements (C vs. Cl and C vs. Br) and compounds. The carbon isotope fractionation was generally more remarkable than the chlorine/bromine isotope fractionation. The mechanistic explanation for the two-dimensional isotope fractionations was tentatively proposed by means of the modified two-film model and the inferences derived from the ZPE theories. The inverse liquid-vapor isotope effects and the normal diffusive isotope effects jointly determined the magnitudes and directions of the observed two-dimensional isotope fractionations, with the former outweighing the latter. The substitution of $^{12}$C with $^{13}$C may lead to larger ZPE differences than the substitution of $^{35}$Cl/$^{79}$Br with $^{37}$Cl/$^{81}$Br for HOCs, which finally result in more significant isotope fractionation of carbon than chlorine/bromine as observed in this study. The findings of this study may provide new insights into the concurrent two-dimensional carbon and chlorine/bromine isotope fractionations in physical changes like dissolution, partition, condensation, vaporization and diffusion. The provided proposals for CISA research using GC are useful to obtain high-quality data, especially for dual-elements CISA studies.

# Appendix A. Supplementary data

The *Supplementary data* is available on the website at http://pending.

# Author information

**Corresponding author**

*Email: CaimingTang@gig.ac.cn (C. Tang).

# Acknowledgements

This study was financially supported by the National Natural Science Foundation of China (Grant No. 41603092).

**Figure legends**

**Figure 1**. Representative chromatograms and high resolution mass spectra of the eight investigated halogenated organic compounds (HOCs). PCB: polychlorinated biphenyl, Me-TCS: methyl-triclosan, MoBB: monobromobenzene, TrBB: 1,3,5-tribromobenzene, PeBB: pentabromobenzene, HBB: hexabromobenzene, NL: nominal level, *m/z*: mass to charge ratio.

**Figure 2**. Chlorine/bromine isotope ratios (IR: $^{37}Cl/^{35}Cl$ or $^{81}Br/^{79}Br$) of the investigated HOCs in three retention-time segments. PCB-18: 2,2'5-trichloro-1,1'biphenyl, PCB-28: 2,4,4'-trichloro-1,1'biphenyl, PCB-52: 2,2',5,5'-tetrachloro-1,1'biphenyl, Me-TCS: methyl-triclosan, MoBB: monobromobenzene, TrBB: 1,3,5-tribromobenzene, PeBB: pentabromobenzene, HBB: hexabromobenzene. T1, T2 and T3 denote the first, the middle and the last retention-time segments, respectively; error bars represent the standard deviations (SD, 1σ, n = 6).

**Figure 3**. Carbon isotope ratios ($^{13}C/^{12}C$) of the investigated HOCs in different retention-time segments.

**Figure 4**. Relative variations of chlorine/bromine and carbon isotope ratios ($\Delta^{37}Cl/\Delta^{81}Br$ and $\Delta^{13}C$) of the investigated HOCs in different retention-time segments. The dashed zero lines indicate the level at $\Delta^{h}E = 0$.

**Figure 5**. Chlorine/bromine and carbon isotope fractionation extents ($\Lambda^{37}Cl/\Lambda^{81}Br$ and $\Lambda^{13}C$) of the investigated HOCs on gas chromatography (GC).

**Figure 6**. $\Delta^{37}Cl$ vs. $\Delta^{13}C$ and $\Delta^{81}Br$ vs. $\Delta^{13}C$ plots of the investigated HOCs for illustrating two-dimensional carbon and chlorine/bromine isotope fractionations on GC. The data of individual compounds were fitted with linear functions. Details of the functions are provided as the following. PCB-18: y = 0.27254x + 0.03235 ($R^2$ = 0.99935), PCB-28: y = 0.27277x + 0.10688 ($R^2$ = 0.99946), PCB-52: y = 0.24151x – 0.32367 ($R^2$ = 0.99867), Me-TCS: y = 0.51708x – 0.12041 ($R^2$ = 0.99712), MoBB: y = 0.58776x – 1.08421 ($R^2$ = 0.69301), TrBB: y = 0.91005x – 0.27609 ($R^2$ = 0.95726), PeBB: y = 0.29374x – 0.27622 ($R^2$ = 0.98083), HBB: y = 0.20494x + 0.24153 ($R^2$ = 0.99823).

# Figures

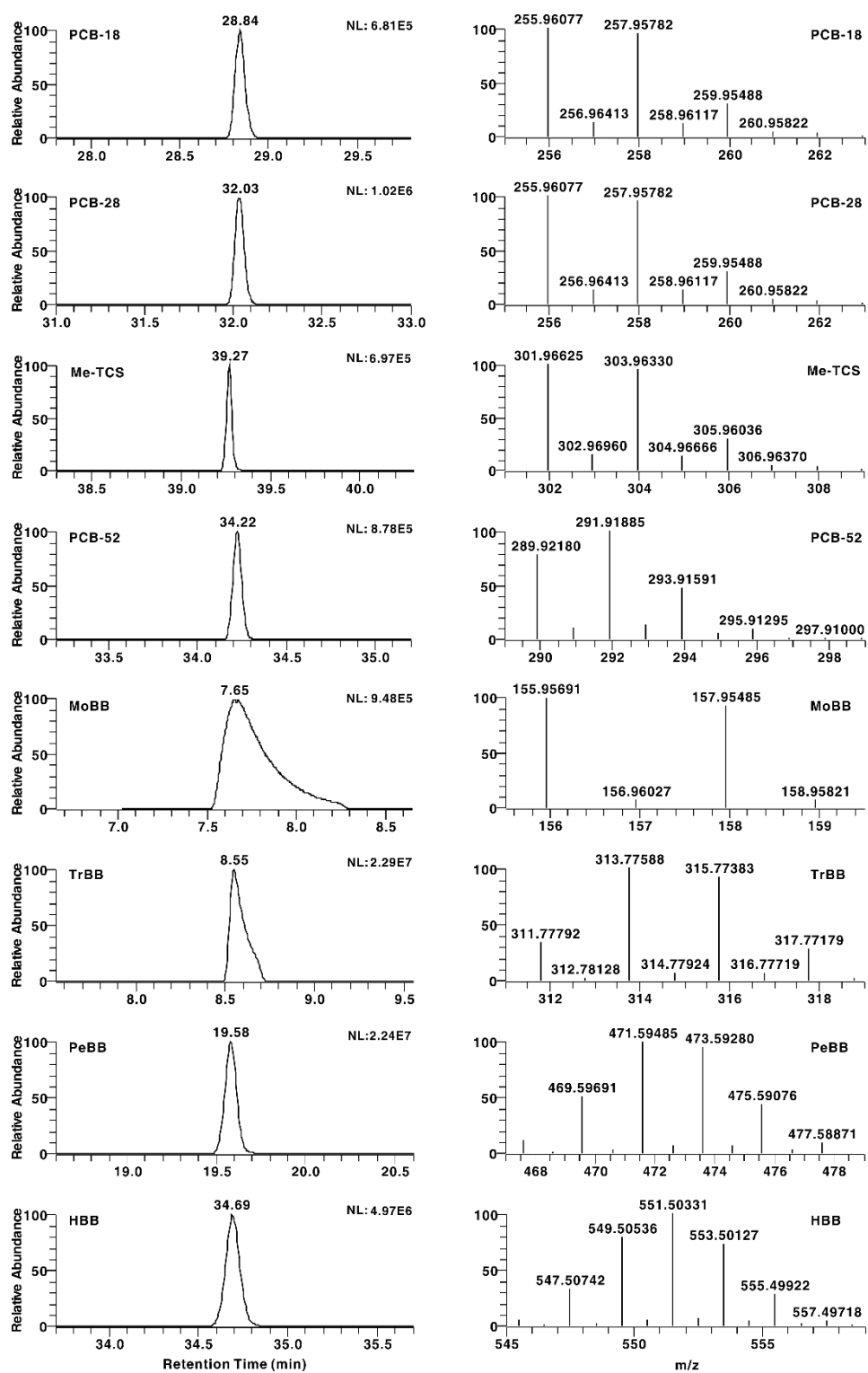

**Figure 1**. Representative chromatograms and high resolution mass spectra of the eight investigated halogenated organic compounds (HOCs). PCB: polychlorinated biphenyl, Me-TCS: methyl-triclosan, MoBB: monobromobenzene, TrBB: 1,3,5-tribromobenzene, PeBB: pentabromobenzene, HBB: hexabromobenzene, NL: nominal level, m/z: mass to charge ratio.

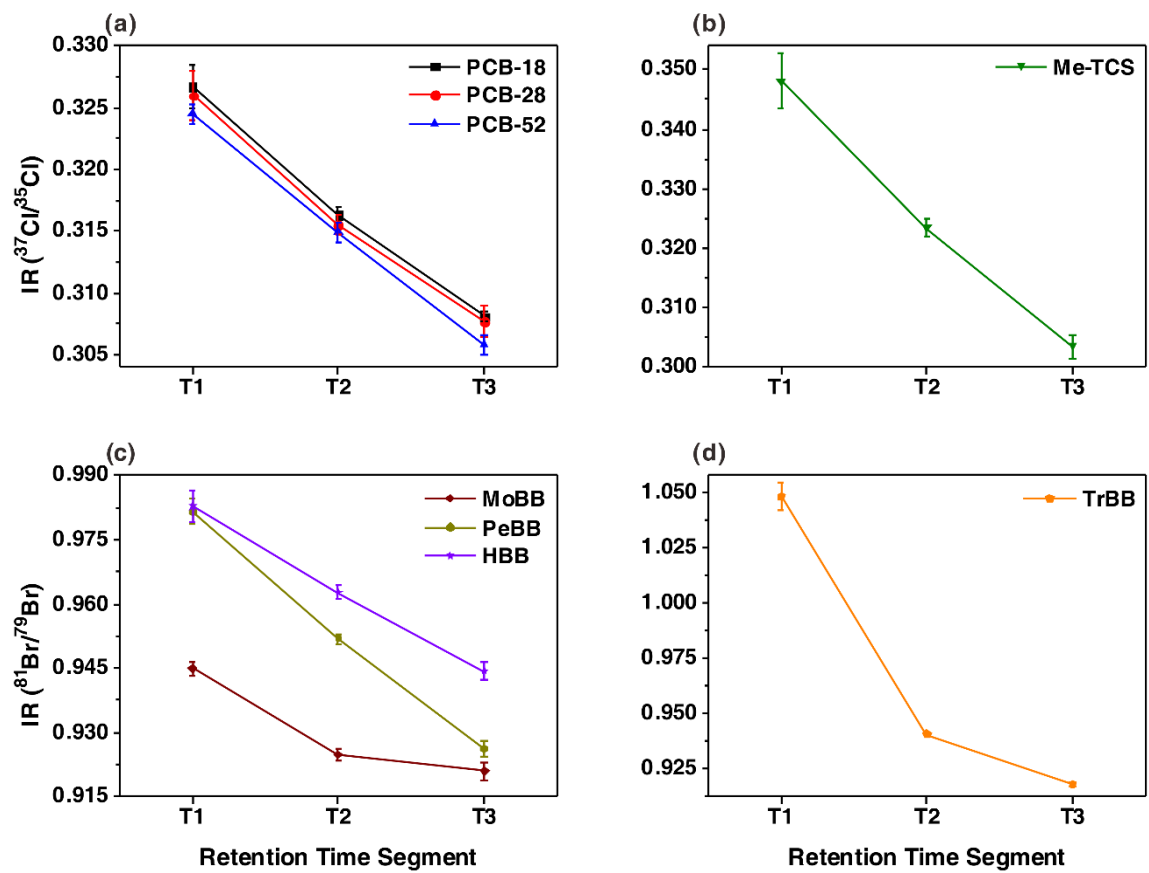

**Figure 2**. Chlorine/bromine isotope ratios (IR: $^{37}Cl/^{35}Cl$ or $^{81}Br/^{79}Br$) of the investigated HOCs in three retention-time segments. PCB-18: 2,2'5-trichloro-1,1'biphenyl, PCB-28: 2,4,4'-trichloro-1,1'biphenyl, PCB-52: 2,2',5,5'-tetrachloro-1,1'biphenyl, Me-TCS: methyl-triclosan, MoBB: monobromobenzene, TrBB: 1,3,5-tribromobenzene, PeBB: pentabromobenzene, HBB: hexabromobenzene. T1, T2 and T3 denote the first, the middle and the last retention-time segments, respectively; error bars represent the standard deviations (SD, 1σ, n = 6).

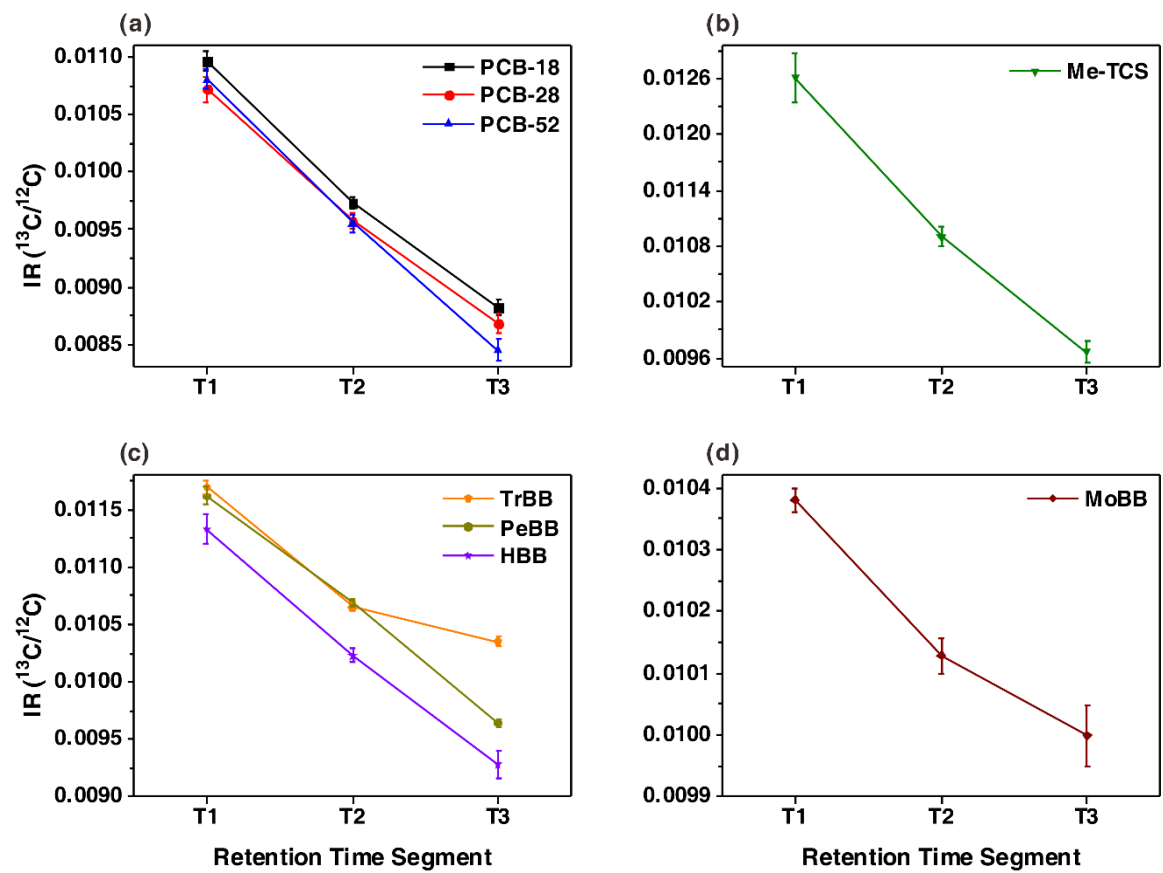

**Figure 3**. Carbon isotope ratios ($^{13}C/^{12}C$) of the investigated HOCs in different retention-time segments.

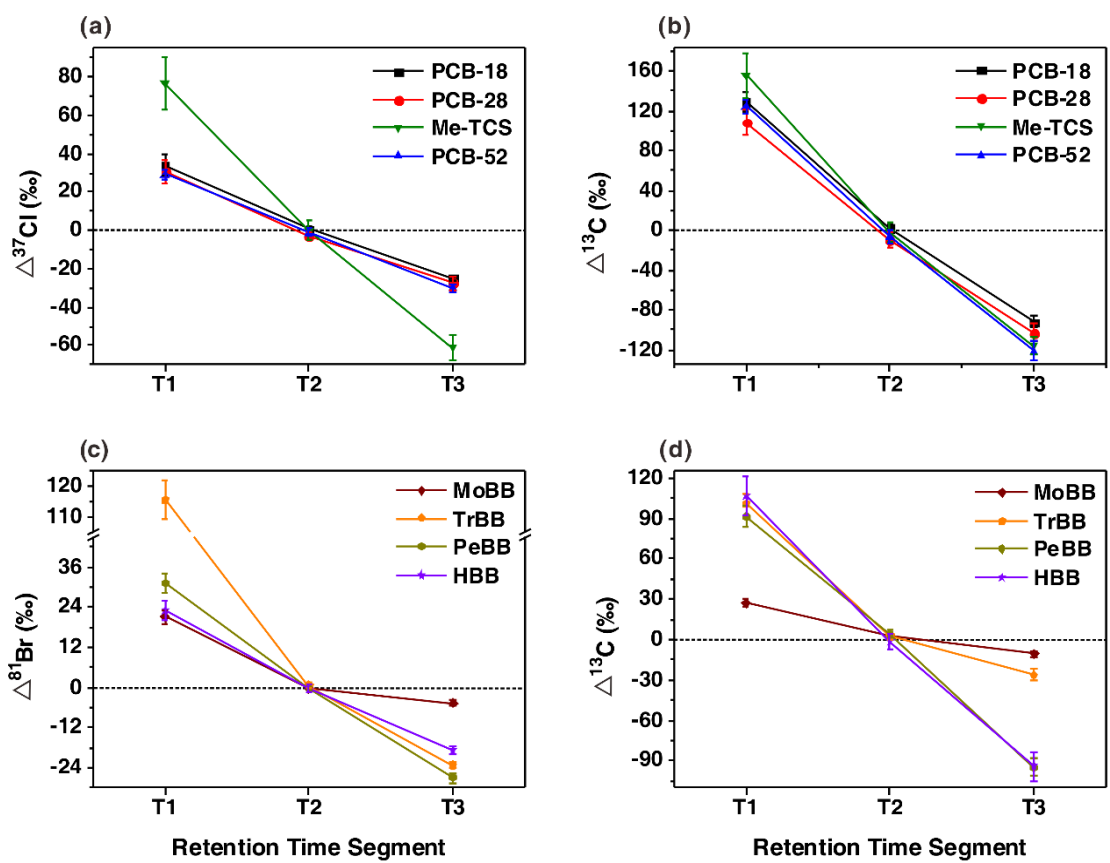

**Figure 4**. Relative variations of chlorine/bromine and carbon isotope ratios ($\Delta^{37}Cl/\Delta^{81}Br$ and $\Delta^{13}C$) of the investigated HOCs in different retention-time segments. The dashed zero lines indicate the level at $\Delta^h E = 0$.

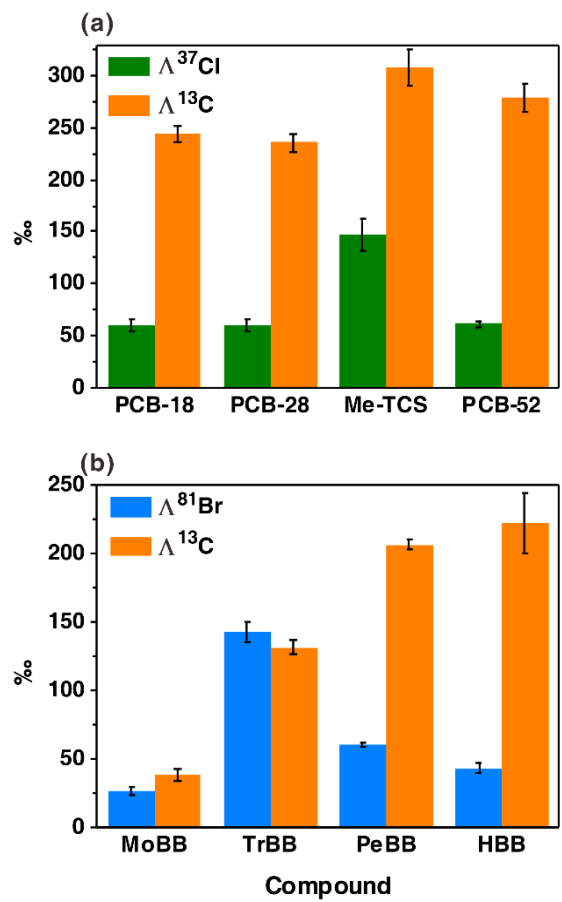

**Figure 5**. Chlorine/bromine and carbon isotope fractionation extents ($\Lambda^{37}Cl/\Lambda^{81}Br$ and $\Lambda^{13}C$) of the investigated HOCs on gas chromatography (GC).

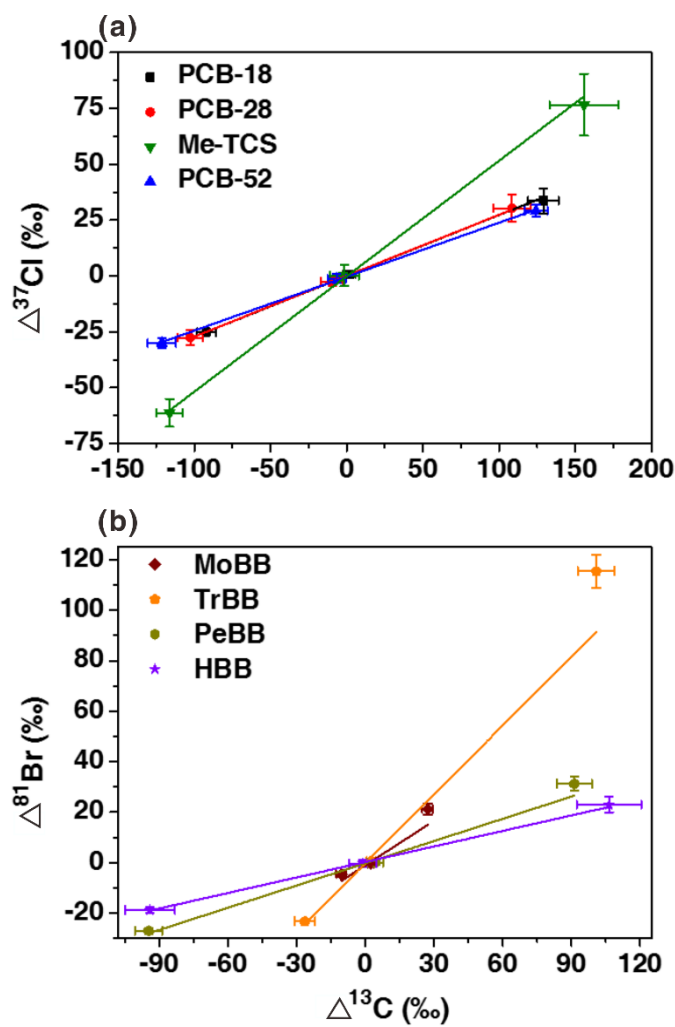

**Figure 6**. $\Delta^{37}$Cl vs. $\Delta^{13}$C and $\Delta^{81}$Br vs. $\Delta^{13}$C plots of the investigated HOCs for illustrating two-dimensional carbon and chlorine/bromine isotope fractionations on GC. The data of individual compounds were fitted with linear functions. Details of the functions are provided as the following. PCB-18: y = 0.27254x + 0.03235 ($R^2$ = 0.99935), PCB-28: y = 0.27277x + 0.10688 ($R^2$ = 0.99946), PCB-52: y = 0.24151x – 0.32367 ($R^2$ = 0.99867), Me-TCS: y = 0.51708x – 0.12041 ($R^2$ = 0.99712), MoBB: y = 0.58776x – 1.08421 ($R^2$ = 0.69301), TrBB: y = 0.91005x – 0.27609 ($R^2$ = 0.95726), PeBB: y = 0.29374x – 0.27622 ($R^2$ = 0.98083), HBB: y = 0.20494x + 0.24153 ($R^2$ = 0.99823).

# SUPPLEMENTARY INFORMATION

# Simultaneous observation of concurrent two-dimensional carbon and chlorine/bromine isotope fractionations of halogenated organic compounds on gas chromatography

Caiming Tang[a,]*, Jianhua Tan[b]

[a] *State Key Laboratory of Organic Geochemistry, Guangzhou Institute of Geochemistry, Chinese Academy of Sciences, Guangzhou 510640, China*

[b] *Guangzhou Quality Supervision and Testing Institute, Guangzhou, 510110, China*

*Corresponding Author.

Tel: +86-020-85291489; Fax: +86-020-85290009; E-mail: CaimingTang@gig.ac.cn (C. Tang).

## Tables

**Table S-1.** Names, structures, CAS numbers, concentrations and chromatographic separation conditions of the investigated halogenated organic compounds (HOCs).

| Compound | Abbreviation | Structure | CAS No. | Temperature program | Concentration (µg/mL) | Injection solvent |
|---|---|---|---|---|---|---|
| 2,2'5-Trichloro-1,1'biphenyl | PCB-18 | 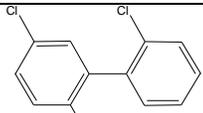 | 37680-65-2 | held at 120 ºC for 2 min, ramped to 220 ºC at 3 ºC/min, held for 2 min, then ramped to 320 ºC at 20 ºC/min, held for 0.67 min. | 1.0 | Isooctane |
| 2,4,4'-Trichloro-1,1'biphenyl | PCB-28 | 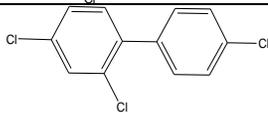 | 7012-37-5 | | 1.0 | Isooctane |
| 2,2',5,5'-Tetrachloro-1,1'biphenyl | PCB-52 | 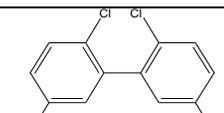 | 35693-99-3 | | 1.0 | Isooctane |
| Methyl-triclosan | Me-TCS | 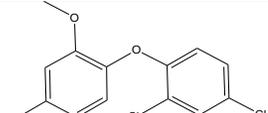 | 4640-01-1 | | 1.0 | Nonane |
| Monobromobenzene | MoBB | 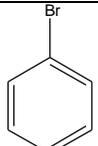 | 108-86-1 | held at 80 ºC for 1 min, ramped to 160 ºC at 5 ºC/min. | 2.0 | Isooctane |
| 1,3,5-Tribromobenzene | TrBB | 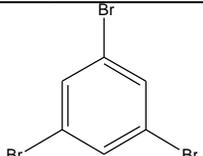 | 626-39-1 | held at 120 ºC for 2 min, ramped to 220 ºC at 20 ºC/min, held for 16 min, then ramped to 235 ºC at 5 ºC/min, held for 7 min, then ramped to 260 ºC | 5.0 | Isooctane |

| Compound | Abbreviation | Structure | CAS No. | Temperature program | Concentration (μg/mL) | Injection solvent |
|---|---|---|---|---|---|---|
| Pentabromobenzene | PeBB | 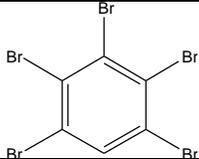 | 608-90-2 | at 5 ºC/min, ramped to 310 ºC at 40 ºC/min, held for 0.75 min. | 5.0 | Isooctane |
| Hexabromobenzene | HBB | 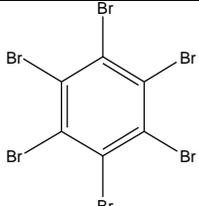 | 87-82-1 | | 5.0 | Isooctane |

**Table S-2.** Retention times, chemical formulas, isotopologue formulas, exact molecular weights and exact *m/z* values of the investigated HOCs.

| Compound | Retention time (min) | Formula | Isotopologue formula | Exact molecular weight (u) | Exact *m/z* value (u) |
|---|---|---|---|---|---|
| PCB-18 | 28.84 | $C_{12}H_7Cl_3$ | $^{12}C_{12}H_7{}^{35}Cl_3$ | 255.96133 | 255.96078 |
| | | | $^{12}C_{11}{}^{13}CH_7{}^{35}Cl_3$ | 256.96468 | 256.96413 |
| | | | $^{12}C_{12}H_7{}^{35}Cl_2{}^{37}Cl$ | 257.95838 | 257.95783 |
| | | | $^{12}C_{11}{}^{13}CH_7{}^{35}Cl_2{}^{37}Cl$ | 258.96173 | 258.96118 |
| | | | $^{12}C_{12}H_7{}^{35}Cl{}^{37}Cl_2$ | 259.95543 | 259.95488 |
| | | | $^{12}C_{11}{}^{13}CH_7{}^{35}Cl{}^{37}Cl_2$ | 260.95878 | 260.95823 |
| | | | $^{12}C_{12}H_7{}^{37}Cl_3$ | 261.95248 | 261.95193 |
| | | | $^{12}C_{11}{}^{13}CH_7{}^{37}Cl_3$ | 262.95583 | 262.95528 |
| | | | | | |
| PCB-28 | 32.03 | $C_{12}H_7Cl_3$ | Refer to PCB-18 | | |
| | | | | | |
| PCB-52 | 34.22 | $C_{12}H_6Cl_4$ | $^{12}C_{12}H_6{}^{35}Cl_4$ | 289.92236 | 289.92181 |
| | | | $^{12}C_{11}{}^{13}CH_6{}^{35}Cl_4$ | 290.92571 | 290.92516 |
| | | | $^{12}C_{12}H_6{}^{35}Cl_3{}^{37}Cl$ | 291.91941 | 291.91886 |
| | | | $^{12}C_{11}{}^{13}CH_6{}^{35}Cl_3{}^{37}Cl$ | 292.92276 | 292.92221 |
| | | | $^{12}C_{12}H_6{}^{35}Cl_2{}^{37}Cl_2$ | 293.91646 | 293.91591 |
| | | | $^{12}C_{11}{}^{13}CH_6{}^{35}Cl_2{}^{37}Cl_2$ | 294.91981 | 294.91926 |
| | | | $^{12}C_{12}H_6{}^{35}Cl{}^{37}Cl_3$ | 295.91351 | 295.91296 |
| | | | $^{12}C_{11}{}^{13}CH_6{}^{35}Cl{}^{37}Cl_3$ | 296.91686 | 296.91631 |
| | | | $^{12}C_{12}H_6{}^{37}Cl_4$ | 297.91056 | 297.91001 |
| | | | $^{12}C_{11}{}^{13}CH_6{}^{37}Cl_4$ | 298.91391 | 298.91336 |
| | | | | | |
| Me-TCS | 39.27 | $C_{13}H_9Cl_3O_2$ | $^{12}C_{13}H_9{}^{35}Cl_3O_2$ | 301.96681 | 301.96626 |
| | | | $^{12}C_{12}{}^{13}CH_9{}^{35}Cl_3O_2$ | 302.97016 | 302.96961 |
| | | | $^{12}C_{13}H_9{}^{35}Cl_2{}^{37}ClO_2$ | 303.96386 | 303.96331 |
| | | | $^{12}C_{12}{}^{13}CH_9{}^{35}Cl_2{}^{37}ClO_2$ | 304.96721 | 304.96666 |
| | | | $^{12}C_{13}H_9{}^{35}Cl{}^{37}Cl_2O_2$ | 305.96091 | 305.96036 |
| | | | $^{12}C_{12}{}^{13}CH_9{}^{35}Cl{}^{37}Cl_2O_2$ | 306.96426 | 306.96371 |
| | | | $^{12}C_{13}H_9{}^{37}Cl_3O_2$ | 307.95796 | 307.95741 |
| | | | $^{12}C_{12}{}^{13}CH_9{}^{37}Cl_3O_2$ | 308.96131 | 308.96076 |
| | | | | | |
| MoBB | 7.65 | $C_6H_5Br$ | $^{12}C_6H_5{}^{79}Br$ | 155.95746 | 155.95691 |
| | | | $^{12}C_5{}^{13}CH_5{}^{79}Br$ | 156.96082 | 156.96027 |
| | | | $^{12}C_6H_5{}^{81}Br$ | 157.95541 | 157.95486 |
| | | | $^{12}C_5{}^{13}CH_5{}^{81}Br$ | 158.95877 | 158.95822 |
| | | | | | |
| TrBB | 8.55 | $C_6H_3Br_3$ | $^{12}C_6H_3{}^{79}Br_3$ | 311.77848 | 311.77793 |
| | | | $^{12}C_5{}^{13}CH_3{}^{79}Br_3$ | 312.78184 | 312.78129 |
| | | | $^{12}C_6H_3{}^{79}Br_2{}^{81}Br$ | 313.77644 | 313.77589 |
| | | | $^{12}C_5{}^{13}CH_3{}^{79}Br_2{}^{81}Br$ | 314.77979 | 314.77925 |
| | | | $^{12}C_6H_3{}^{79}Br{}^{81}Br_2$ | 315.77439 | 315.77384 |
| | | | $^{12}C_5{}^{13}CH_3{}^{79}Br{}^{81}Br_2$ | 316.77774 | 316.77720 |
| | | | $^{12}C_6H_3{}^{81}Br_3$ | 317.77234 | 317.77179 |

| Compound | Retention time (min) | Formula | Isotopologue formula | Exact molecular weight (u) | Exact m/z value (u) |
|---|---|---|---|---|---|
| | | | $^{12}C_5^{13}CH_3^{81}Br_3$ | 318.77570 | 318.77515 |
| | | | | | |
| PeBB | 19.58 | $C_6HBr_5$ | $^{12}C_6H^{79}Br_5$ | 467.59950 | 467.59895 |
| | | | $^{12}C_5^{13}CH^{79}Br_5$ | 468.60285 | 468.60231 |
| | | | $^{12}C_6H^{79}Br_4^{81}Br$ | 469.59746 | 469.59691 |
| | | | $^{12}C_5^{13}CH^{79}Br_4^{81}Br$ | 470.60081 | 470.60027 |
| | | | $^{12}C_6H^{79}Br_3^{81}Br_2$ | 471.59541 | 471.59486 |
| | | | $^{12}C_5^{13}CH^{79}Br_3^{81}Br_2$ | 472.59877 | 472.59822 |
| | | | $^{12}C_6H^{79}Br_2^{81}Br_3$ | 473.59336 | 473.59281 |
| | | | $^{12}C_5^{13}CH^{79}Br_2^{81}Br_3$ | 473.59336 | 474.59617 |
| | | | $^{12}C_6H^{79}Br^{81}Br_4$ | 475.59132 | 475.59077 |
| | | | $^{12}C_5^{13}CH^{79}Br^{81}Br_4$ | 476.59467 | 476.59413 |
| | | | $^{12}C_6H^{81}Br_5$ | 477.58927 | 477.58872 |
| | | | $^{12}C_5^{13}CH^{81}Br_5$ | 478.59262 | 478.59208 |
| | | | | | |
| HBB | 34.69 | $C_6Br_6$ | $^{12}C_6^{79}Br_6$ | 545.51002 | 545.50947 |
| | | | $^{12}C_5^{13}C^{79}Br_6$ | 546.51337 | 546.51283 |
| | | | $^{12}C_6^{79}Br_5^{81}Br$ | 547.50797 | 547.50742 |
| | | | $^{12}C_5^{13}C^{79}Br_5^{81}Br$ | 548.51132 | 548.51078 |
| | | | $^{12}C_6^{79}Br_4^{81}Br_2$ | 549.50592 | 549.50537 |
| | | | $^{12}C_5^{13}C^{79}Br_4^{81}Br_2$ | 550.50928 | 550.50873 |
| | | | $^{12}C_6^{79}Br_3^{81}Br_3$ | 551.50387 | 551.50332 |
| | | | $^{12}C_5^{13}C^{79}Br_3^{81}Br_3$ | 552.50723 | 552.50668 |
| | | | $^{12}C_6^{79}Br_2^{81}Br_4$ | 553.50183 | 553.50128 |
| | | | $^{12}C_5^{13}C^{79}Br_2^{81}Br_4$ | 554.50518 | 554.50464 |
| | | | $^{12}C_6^{79}Br^{81}Br_5$ | 555.49978 | 555.49923 |
| | | | $^{12}C_5^{13}C^{79}Br^{81}Br_5$ | 556.50314 | 556.50259 |
| | | | $^{12}C_6^{81}Br_6$ | 557.49773 | 557.49718 |
| | | | $^{12}C_5^{13}C^{81}Br_6$ | 558.50108 | 558.50054 |

**Table S-3.** Overall isotope ratios (IR$_{overall}$) and precisions (standard deviations) of all the investigated HOCs.

| Compound | Chlorine/bromine isotope ratio (mean, n=6) | SD (1σ, n=6, ‰) | Carbon isotope ratio (mean, n=6) | SD (1σ, n=6, ‰) |
|---|---|---|---|---|
| PCB-18 | 0.31605 | 0.40 | 0.00971 | 0.04 |
| PCB-28 | 0.31640 | 0.41 | 0.00967 | 0.02 |
| PCB-52 | 0.31527 | 0.32 | 0.00961 | 0.04 |
| Me-TCS | 0.32336 | 0.32 | 0.01092 | 0.05 |
| MoBB | 0.92525 | 1.22 | 0.01011 | 0.03 |
| TrBB | 0.93949 | 0.94 | 0.01062 | 0.02 |
| PeBB | 0.95188 | 0.91 | 0.01065 | 0.01 |
| HBB | 0.96303 | 1.30 | 0.01024 | 0.04 |

Note, SD: Standard deviation.

**Table S-4.** Isotope ratios, relative variations of isotope ratios derived from different retention-time segments (Δ$^h$E) and isotope fractionation extents (Λ$^h$E) of the investigated HOPs.

| Compound | Retention-time segment | Cl/Br isotope ratio (mean, n=6) | SD (1σ, n=6, ‰) | Δ$^{37}$Cl/Δ$^{81}$Br (mean, n=6, ‰) | SD (1σ, n=6, ‰) | Λ$^{37}$Cl/Λ$^{81}$Br (mean, n=6, ‰) | SD (1σ, n=6, ‰) | C isotope ratio (mean, n=6) | SD (1σ, n=6, ‰) | Δ$^{13}$C (mean, n=6, ‰) | SD (1σ, n=6, ‰) | Λ$^{13}$C (mean, n=6, ‰) | SD (1σ, n=6, ‰) |
|---|---|---|---|---|---|---|---|---|---|---|---|---|---|
| PCB-18 | T1 | 0.32671 | 1.80 | 33.71 | 5.65 | 60.34 | 5.79 | 0.01097 | 0.09 | 128.98 | 10.11 | 243.48 | 7.81 |
|  | T2 | 0.31630 | 0.63 | 0.78 | 1.51 |  |  | 0.00973 | 0.05 | 1.82 | 4.04 |  |  |
|  | T3 | 0.30812 | 0.39 | -25.11 | 0.88 |  |  | 0.00882 | 0.07 | -92.08 | 6.19 |  |  |
| PCB-28 | T1 | 0.32599 | 2.04 | 30.31 | 5.89 | 59.60 | 5.22 | 0.01072 | 0.11 | 108.41 | 12.36 | 235.15 | 9.32 |
|  | T2 | 0.31557 | 0.80 | -2.61 | 1.88 |  |  | 0.00958 | 0.07 | -9.43 | 7.37 |  |  |
|  | T3 | 0.30765 | 1.29 | -27.64 | 3.24 |  |  | 0.00868 | 0.08 | -102.60 | 8.49 |  |  |
| PCB-52 | T1 | 0.32446 | 0.75 | 29.15 | 2.60 | 60.98 | 2.84 | 0.01081 | 0.08 | 124.25 | 7.40 | 279.49 | 13.52 |
|  | T2 | 0.31492 | 0.83 | -1.11 | 2.15 |  |  | 0.00955 | 0.08 | -6.51 | 6.04 |  |  |
|  | T3 | 0.30581 | 0.78 | -30.00 | 2.09 |  |  | 0.00845 | 0.09 | -121.26 | 9.25 |  |  |
| Me-TCS | T1 | 0.34812 | 4.58 | 76.57 | 13.83 | 146.85 | 14.70 | 0.01261 | 0.27 | 155.68 | 22.66 | 307.56 | 17.92 |
|  | T2 | 0.32342 | 1.52 | 0.19 | 4.80 |  |  | 0.01090 | 0.11 | -1.51 | 9.69 |  |  |
|  | T3 | 0.30355 | 2.00 | -61.27 | 6.24 |  |  | 0.00966 | 0.12 | -116.19 | 8.57 |  |  |
| MoBB | T1 | 0.94482 | 1.60 | 21.15 | 2.20 | 25.89 | 2.97 | 0.01038 | 0.02 | 27.25 | 2.39 | 38.14 | 4.00 |
|  | T2 | 0.92482 | 1.27 | -0.47 | 0.44 |  |  | 0.01013 | 0.03 | 2.28 | 1.31 |  |  |
|  | T3 | 0.92098 | 1.96 | -4.62 | 1.02 |  |  | 0.01000 | 0.05 | -10.49 | 2.09 |  |  |
| TrBB | T1 | 1.04828 | 6.19 | 115.51 | 6.54 | 142.10 | 6.96 | 0.01170 | 0.06 | 100.85 | 7.95 | 130.72 | 4.95 |
|  | T2 | 0.94033 | 0.88 | 0.64 | 0.76 |  |  | 0.01065 | 0.04 | 2.57 | 1.94 |  |  |
|  | T3 | 0.91760 | 1.11 | -23.29 | 0.74 |  |  | 0.01035 | 0.04 | -26.43 | 4.46 |  |  |
| PeBB | T1 | 0.98158 | 2.89 | 31.21 | 2.80 | 59.95 | 2.04 | 0.01162 | 0.08 | 91.37 | 7.66 | 205.39 | 3.70 |
|  | T2 | 0.95184 | 1.18 | -0.05 | 0.97 |  |  | 0.01069 | 0.03 | 4.34 | 3.42 |  |  |

| Compound | Retention-time segment | Cl/Br isotope ratio (mean, n=6) | SD (1σ, n=6, ‰) | $\Delta^{37}Cl/\Delta^{81}Br$ (mean, n=6, ‰) | SD (1σ, n=6, ‰) | $\Lambda^{37}Cl/\Lambda^{81}Br$ (mean, n=6, ‰) | SD (1σ, n=6, ‰) | C isotope ratio (mean, n=6) | SD (1σ, n=6, ‰) | $\Delta^{13}C$ (mean, n=6, ‰) | SD (1σ, n=6, ‰) | $\Lambda^{13}C$ (mean, n=6, ‰) | SD (1σ, n=6, ‰) |
|---|---|---|---|---|---|---|---|---|---|---|---|---|---|
| | T3 | 0.92607 | 1.76 | -27.12 | 1.35 | | | 0.00964 | 0.03 | -94.59 | 5.96 | | |
| HBB | T1 | 0.98292 | 3.70 | 22.97 | 3.06 | 42.62 | 3.45 | 0.01133 | 0.13 | 106.45 | 14.04 | 221.45 | 22.31 |
| | T2 | 0.96274 | 1.69 | -0.32 | 1.27 | | | 0.01023 | 0.06 | -1.09 | 5.90 | | |
| | T3 | 0.94436 | 2.08 | -18.84 | 1.30 | | | 0.00928 | 0.12 | -94.03 | 10.85 | | |